\documentclass[physrev]{revtex4-2}

\usepackage{graphicx}
\usepackage{dcolumn}
\usepackage{bm}
\usepackage{centernot}
\usepackage{amsmath}
\usepackage{amssymb}

\usepackage{enumitem}
\usepackage[normalem]{ulem}

\usepackage{hyperref}
\hypersetup{colorlinks,breaklinks,
			citecolor=[rgb]{0,0.0,1.0},
            urlcolor=[rgb]{0.0,0.0,1.0},
            linkcolor=[rgb]{0,0.5,0.9}}

\begin{document}

\title{\Large \textbf{On energy-momentum conservation in non-minimal geometry-matter coupling theories} 
}

\author{Gonzalo J. Olmo}
\email{gonzalo.olmo@uv.es}
\affiliation{\small Instituto de Física Corpuscular (IFIC), CSIC‐Universitat de València, Spain}
\affiliation{\small Universidade Federal do Cear\'a (UFC), Departamento de F\'isica,\\ Campus do Pici, Fortaleza - CE, C.P. 6030, 60455-760 - Brazil}

\author{Miguel A. S. Pinto}
\email{mapinto@fc.ul.pt}
\affiliation{\small Instituto de Astrof\'{i}sica e Ci\^{e}ncias do Espa\c{c}o, Faculdade de Ci\^{e}ncias da Universidade de Lisboa, Edif\'{i}cio C8, Campo Grande, P-1749-016 Lisbon, Portugal.}
\affiliation{\small Departamento de F\'{i}sica, Faculdade de Ci\^{e}ncias da Universidade de Lisboa,  Edif\'{i}cio C8, Campo Grande, P-1749-016 Lisbon, Portugal.}

\date{\today}

\begin{abstract}
In this work, we discuss the conditions that allow the establishment of an equivalence between $f(R,T)=R+\lambda h(T)$ gravity models and General Relativity (GR) coupled to a modified matter sector. We do so by considering a $D$-dimensional spacetime and the matter sector to be described by nonlinear electrodynamics and/or a scalar field. We find that, for this particular family of models, the action and field equations can indeed be written in terms of a modified matter source within GR. However, when several matter sources are combined, this interpretation is no longer possible if $h(T)$ is a nonlinear function, due to the emergence of crossed terms that mix together the scalar and vector sectors.
\end{abstract}

\keywords{geometry-matter couplings, modified gravity, nonlinear electrodynamics, scalar field}

\maketitle
\thispagestyle{empty}


\setcounter{page}{1} 

\section{Introduction}
Despite the vast amount of experimental and observational data that support General Relativity (GR) as our standard gravitational theory in local and astrophysical scenarios, the strongly sustained finding that the expansion of the universe is currently accelerating \cite{acc1,acc2,acc3,acc4,acc5,Planck:2018vyg} has led the international community to question whether GR, through the standard $\Lambda$ Cold Dark Matter ($\Lambda$CDM) model, is a sufficiently good description of gravitational phenomena at large cosmological scales. In fact, cosmology poses an ideal playground to test GR, as more observations are suggesting inconsistencies within the standard cosmological model, with the Hubble tension being the most notable one \cite{CosmoVerseNetwork:2025alb,DiValentino:2021izs}. In this regard, many generalizations of and alternatives to the Einstein-Hilbert action have been formulated and have become collectively known as modified theories of gravity \cite{MG1,MG2,MG3}. Indeed, modified gravity models propose, among other aspects, alternative descriptions of the late-time cosmic acceleration that rely on extra fields or geometrical contributions rather than on the cosmological constant, which is the main characteristic of the $\Lambda$CDM model.

One specific class of modified theories of gravity assumes that the matter and geometry sectors can be non-minimally coupled to each other; therefore, they are often referred to as non-minimal geometry-matter coupling theories \cite{Harko:2010mv,Harko:2011kv,Katirci:2013okf,Haghani:2013oma,Harko:2014aja,Harko:2014sja,Harko:2018gxr,Harko:2021bdi}. In this context, a non-minimal coupling between the two sectors means that at the action level, the geometry and matter sectors are no longer separated, as in GR (and, more generally, in all metric theories of gravity).  In such theoretical frameworks, a fifth force can appear, offering alternative mechanisms to justify the anomalous dynamics of rotation curves in spiral galaxies \cite{Bertolami:2007gv}. Moreover, the usual energy-momentum tensor of matter is generally non-conserved, leading to violations of the equivalence principle. This non-conservation has been interpreted in the literature mainly as two-sided. While one interpretation assumes that the geometry sector exchanges energy and momentum with the matter sector, yielding a phenomenological production of macroscopic particles according to the irreversible thermodynamics of open systems \cite{Harko:2014pqa,Pinto:2023phl}, the other states that the momenta of particles are not evolving as demanded by momentum and energy conservation, but differently \cite{Azevedo:2019oah,Avelino:2022eqm}. 

Among the criticisms that non-minimal geometry-matter couplings have faced throughout the years, one of them is of particular interest. To illustrate it, let us consider the so-called $f(R,T)$ gravity theory \cite{Harko:2011kv}, where $R$ is the Ricci scalar and $T$ is the trace of the energy-momentum tensor of matter $T_{\mu\nu}$. It has been argued that when the function has the form $f(R,T)=R + \lambda h(T)$, where $h(T)$ is an analytic function of $T$, then the theory is not a genuine modification of the gravitational sector. In other words, despite having a general non-conservation of the energy-momentum tensor of matter, the theory does not present an explicit non-minimal coupling between geometry and matter because in such a case one can absorb the function $h(T)$ into the matter sector, hence defining a new, modified matter Lagrangian \cite{Fisher:2019ekh,Lacombe:2023pmx}. Moreover, it has been proven that for other non-minimal geometry-matter coupling theories, models of a similar type are dynamically equivalent to GR with non-minimal matter interactions \cite{Akarsu:2023lre}.

In this paper, we aim to fill a gap in the literature by demonstrating that indeed $f(R,T)$ gravity models in which the function $f$ is given by $f(R,T) = R +\lambda h\left(T\right)$, with $h(T)$ being a function that solely depends on $T$, are physically equivalent to GR plus a modified matter sector, by considering explicit kinds of matter described by nonlinear electrodynamics (NED) and a scalar field. As we will see, when the matter sector is limited to a single source, the resulting field equations are equivalent to those of GR coupled to a nonlinear version of the original matter source. However, if several matter sources are considered together and the function $h(T)$ is nonlinear on $T$, then the coupling between the scalar and vector degrees of freedom prevents a clear splitting of both sectors. In any case, it is always possible to interpret the dynamics as corresponding to GR with a conserved effective energy-momentum tensor.

The paper is organized as follows. In Sec. \ref{secII}, we present the field equations of $f(R,T)$ gravity in their general form. In Sec. \ref{secIII}, we perform our formal analysis by considering a $f(R,T) = R + \lambda h(T)$ gravity model coupled to NED, first particularizing $h(T)=T$ and then taking the general case. In Sec. \ref{secIV}, we perform the very same analysis, but this time assuming $h(T) = T$ only, and that the matter sector is solely constituted by a quintessence scalar type of matter. In Sec. \ref{secV}, we consider that the matter sector is composed of both NED and scalar matter. Finally, we present our main conclusions in Sec. \ref{secVI}.
\section{Field Equations of $f(R,T)$ Gravity}
\label{secII}
The general action of $D$-dimensional $f(R,T)$ gravity is given by 
\begin{equation}
    S=\frac{1}{2 \kappa^2}\int_{\Omega} \sqrt{-g} \, f\left(R,T\right) \, d^Dx + \int_{\Omega} \sqrt{-g} \, \mathcal{L}_m \, d^Dx \, ,
\end{equation}
where $\Omega$ is the $D$-dimensional manifold on which one defines a set of coordinates $x^{\mu}$, $\kappa^2$ is the effective gravitational constant, $g$ is the determinant of the metric $g_{\mu \nu}$, $R$ is the Ricci scalar, and $T$ is the trace of the energy-momentum tensor of matter $T_{\mu\nu}$, which is defined in terms of the matter Lagrangian $\mathcal{L}_m$ by convention as
\begin{equation}
    T_{\mu \nu} := -\frac{2}{\sqrt{-g}} \frac{\delta\left(\sqrt{-g} \mathcal{L}_m\right)}{\delta g^{\mu \nu}} \, .
    \label{main_action}
\end{equation}
The corresponding field equations are obtained by varying Eq. \eqref{main_action} with respect to the metric field
\begin{equation}
\label{eq:frt_campo_geometrical}
f_{R}R_{\mu \nu}-\frac{1}{2} g_{\mu \nu} f+\left(g_{\mu \nu}\square -\nabla_{\mu} \nabla_{\nu}\right) f_{R}=\kappa^{2} T_{\mu \nu}-f_{T}\left(T_{\mu \nu}+\Theta_{\mu \nu}\right)\, ,
\end{equation}
where $f_R$ and $f_T$ denote, respectively, the derivative of $f$ with respect to $R$ and $T$, $\square$ is the D'Alembert opera, defined as $\square := \nabla_{\mu} \nabla^{\mu}$, with $\nabla_\mu$ being the covariant derivative of the Levi-Civita connection, and where we have introduced an auxiliary tensor, $\Theta_{\mu \nu}$, defined as
\begin{equation}
    \Theta_{\mu \nu} := g^{\alpha \beta} \frac{\delta T_{\alpha \beta}}{\delta g^{\mu \nu}} \, .
\end{equation}

Note that we are working in $D$ dimensions to reinforce the generality of our analysis and results, which are by no means restricted to $4$ spacetime dimensions.
\section{$f(R,T) =  R + \lambda h(T)$ Model Coupled to NED }
\label{secIII}

\subsection{Linear Case: $h(T) = T$}
Assuming that the function $f(R,T)$ has the form $f(R,T) = R + \lambda T$, where $\lambda$ is a real constant, and that the matter sector consists of a general nonlinear electrodynamics content, the action \eqref{main_action} becomes 
\begin{equation}
    S=\frac{1}{2 \kappa^2}\int_{\Omega} \sqrt{-g} \, \left(R+\lambda T\right) \, d^Dx + \frac{1}{8\pi}\int_{\Omega} \sqrt{-g} \, \varphi\left(X\right) \, d^Dx \, ,
    \label{action2}
\end{equation}
where we have assumed $\mathcal{L}_m=\mathcal{L}^{\text{NED}} = \frac{\varphi(X)}{8\pi}$, with $\varphi(X)$ being a function of the scalar $X:=-\frac{1}{2} F_{\mu\nu}F^{\mu\nu}$, where $F_{\mu \nu}$ is the Maxwell-Faraday tensor, which is then defined in terms of a vector field, $A_{\mu}$, as $F_{\mu\nu}:=\partial_\mu A_\nu - \partial_\nu A_\mu$. By varying Eq. \eqref{action2} with respect to the metric, we obtain the following field equations
\begin{equation}
    G_{\mu\nu} + \lambda\left(\frac{\delta T^{\text{NED}}}{\delta g^{\mu\nu}}-\frac{1}{2}g_{\mu\nu}T^{\text{NED}}\right) = \kappa^2 T_{\mu\nu}^{\text{NED}} \, ,
    \label{field_eqs_raw}
\end{equation}
where the NED energy-momentum tensor $T_{\mu\nu}^{\text{NED}}$ is given by
\begin{equation}
T_{\mu\nu}^{\text{NED}} = \frac{1}{4\pi}\left(\varphi_X F_{\nu\alpha} F_{\mu}{^{\alpha}} +\frac{1}{2}g_{\mu\nu} \varphi\right) \, .
    \label{energy-momentum tensor NED1}
\end{equation}
The conservation of this energy-momentum tensor can be easily proven in the case $\lambda=0$ by using the equations of motion $\nabla_\mu \left(\varphi_X {F^\mu}_\nu\right)=0$ and the Bianchi identities $\nabla_{[\nu}F_{\alpha\beta]}=0$ to write $F^{\alpha\beta}\nabla_\nu F_{\alpha\beta}=2F^{\mu\alpha}\nabla_\mu F_{\nu\alpha}$. By direct calculation, one finds that 
\begin{equation}\label{eq:cons_Tmn}
4\pi \nabla_\mu {T^\mu}_{\nu}=\nabla_\mu \left(\varphi_X {F^\mu}_\alpha\right){F_\nu}^\alpha+\varphi_X {F^\mu}_\alpha\nabla_\mu {F_\nu}^\alpha+\frac{1}{2}\varphi_X \nabla_\nu X \ ,
\end{equation}
and using that $\nabla_\nu X=-F^{\alpha\beta}\nabla_\nu F_{\alpha\beta}=-2F^{\mu\alpha}\nabla_\mu F_{\nu\alpha}$ and the above tips, one confirms that $\nabla_\mu{T^\mu}_{\nu}=0$. When $\lambda\neq 0$, one must find out how the $\lambda-$dependent terms modify the equations of motion and the divergence of Eq. (\ref{field_eqs_raw}) to check if anything is conserved and how. To proceed, we note that the trace of $T_{\mu\nu}^{\text{NED}}$ is given by
\begin{equation}
    T^{\text{NED}}=\frac{1}{2\pi}\left(\frac{D}{4}\varphi - X\varphi_X \right) \, ,
    \label{NED_trace}
\end{equation}
and 
\begin{equation}
    \frac{\delta T^{\text{NED}}}{\delta g^{\mu\nu}} = \frac{1}{2\pi}\left(X\varphi_{XX}-\frac{D-4}{4}\varphi_X\right)F_{\nu\alpha} F_{\mu}{^{\alpha}} \, .
    \label{NED_trace_variation}
\end{equation}
Therefore, we have the following explicit form for the field equations
\begin{equation}
    G_{\mu\nu} + \frac{\lambda}{2\pi}\left[\left(X \varphi_{XX} -\frac{D-4}{4}\varphi_X\right)F_{\nu\alpha}F_\mu{^{\alpha}} + \frac{1}{2}\left(X\varphi_X-\frac{D}{4}\varphi\right) g_{\mu\nu}\right] = \frac{\kappa^2}{4\pi}\left(\varphi_X F_{\nu \alpha}F_{\mu}{^{\alpha}} +\frac{1}{2} g_{\mu\nu}\varphi\right) \, .   \label{fieldeqsfull}
\end{equation}
Now, notice that the second term on the left-hand side of Eq. \eqref{fieldeqsfull} can be written as a tensor $\tau_{\mu\nu}^{\text{NED}}$ of the form
\begin{equation}
\tau_{\mu\nu}^{\text{NED}} = \frac{\lambda}{2\pi}\left[\left(X\varphi_{XX} - \frac{D-4}{4} \varphi_X\right) F_{\nu \alpha} F_\mu {^{\alpha}} + \frac{1}{2} \left(X \varphi_X -\frac{D}{4}\varphi\right)g_{\mu\nu}\right] \, ,
\label{effective_tensor_1}
\end{equation}
which has the same formal structure as the energy-momentum tensor present in Eq.\eqref{energy-momentum tensor NED1}, in the sense that it can be decomposed into
\begin{equation}
\tau_{\mu\nu}^{\text{NED}} = B\left(X\right)F_{\nu \alpha} F_\mu {^{\alpha}} +\frac{1}{2}g_{\mu\nu} A\left(X\right) \, ,
\label{effective_tensor_2}
\end{equation}
where $B\left(X\right)$ and $A\left(X\right)$ are functions of the scalar $X$ previously defined. By comparing Eqs. \eqref{effective_tensor_1} and \eqref{effective_tensor_2}, one can easily verify that
\begin{equation}
    A\left(X\right) = \frac{\lambda}{2\pi} \left(X \varphi_X -\frac{D}{4}\varphi\right) \, ,
\end{equation}
and, as a consequence, that
\begin{equation}
B\left(X\right)=A_X = \frac{\lambda}{2\pi}\left(X\varphi_{XX}-\frac{D-4}{4}\varphi_X\right)\, .
\end{equation}
Therefore, we can rewrite the field equations as
\begin{equation}
    G_{\mu \nu } = \frac{\kappa^2}{4 \pi} \left[\left(\varphi_X-\frac{4\pi}{\kappa^2}A_X\right)F_{\nu \alpha} F_\mu {^{\alpha}}+\frac{1}{2}g_{\mu\nu}\left(\varphi-\frac{4\pi}{\kappa^2}A\right)\right] \, .
    \label{field_eqs_2}
\end{equation}
If we now define a new scalar function $\Tilde{\varphi}$ as
\begin{equation}
\Tilde{\varphi} := \varphi-\frac{4\pi}{\kappa^2}A=\varphi -\frac{2\lambda}{\kappa^2}\left(X \varphi_X - \frac{D}{4}\varphi \right) \, ,
\label{redef}
\end{equation}
it becomes clear that Eq. \eqref{field_eqs_2} can be directly recovered through the variation with respect to the metric of the following action 
\begin{equation}
    S= \frac{1}{2 \kappa^2}\int_\Omega \sqrt{-g} \, R \, d^Dx + \frac{1}{8\pi}\int_{\Omega}\sqrt{-g}\, \Tilde{\varphi}\left(X\right) \, d^Dx \, ,
\end{equation}
where the energy-momentum tensor of the redefined NED, $\Tilde{T}_{\mu\nu}^{\text{NED}}$, assumes the form
\begin{equation} \Tilde{T}_{\mu\nu}^{\text{NED}}= \frac{1}{4\pi}\left(\Tilde{\varphi}_X F_{\nu\alpha} F_{\mu}{^{\alpha}} +\frac{1}{2}g_{\mu\nu} \Tilde{\varphi}\right) \, ,
\end{equation}
which precisely corresponds to the right-hand side of Eq. \eqref{field_eqs_2}. Therefore, we conclude that an $f(R,T)=R + \lambda T$ gravity model coupled to NED can be interpreted as GR coupled to some modified NED, being its energy-momentum tensor conserved by the same formal manipulations that led to (\ref{eq:cons_Tmn}).

\subsection{General Case}
The previous analysis can also be applied in the case of a $f(R,T) = R + \lambda h(T)$ gravity model coupled to nonlinear electrodynamics, where $h(T)$ is a well-behaved function of the trace of the energy-momentum tensor. In this case, the action is given by
 \begin{equation}
    S = \frac{1}{2 \kappa^2} \int_\Omega\sqrt{-g}\, \left[R + \lambda h(T)\right] \, d^D x + \frac{1}{8\pi}\int_{\Omega} \sqrt{-g} \, \varphi\left(X\right) \, d^Dx \, .
    \label{action3}
 \end{equation}
By varying it with respect to the metric field, the modified field equations are then obtained
\begin{equation}
    G_{\mu \nu} + \lambda \left(h_T \frac{\delta T^{\text{NED}}}{\delta g^{\mu\nu}} -\frac{1}{2}g_{\mu\nu}h\right) = \kappa^2 T_{\mu\nu}^{\text{NED}} \, ,
\end{equation}
where we have denoted the derivative of $h$ with respect to $T$ as $h_T$. Since the matter Lagrangian is the same as in the previous case, $T_{\mu\nu}^{\text{NED}}$ and $\delta T^{\text{NED}}/\delta g^{\mu\nu}$ are again of the form of Eqs. \eqref{energy-momentum tensor NED1} and \eqref{NED_trace_variation}, respectively. As such, the explicit form of the field equations will be just a slightly different version of  Eq. \eqref{fieldeqsfull}
\begin{equation}
    G_{\mu\nu} + \frac{\lambda}{2\pi}\left[h_T\left(X \varphi_{XX} -\frac{D-4}{4}\varphi_X\right)F_{\nu\alpha}F_\mu{^{\alpha}} - \pi g_{\mu\nu} h\right] = \frac{\kappa^2}{4\pi}\left(\varphi_X F_{\nu \alpha}F_{\mu}{^{\alpha}} +\frac{1}{2} g_{\mu\nu}\varphi\right) \, ,   \label{fieldeqsfull}
\end{equation}
with the $h(T)$ function possibly introducing a non-linearity with respect to $T$ into the gravitational dynamics. Following the previous analysis of the linear $h(T)$ theory, we notice that the second term of the left-hand side of Eq. \eqref{fieldeqsfull} has the structure
\begin{equation}
    \tau_{\mu \nu}^{\text{NED}}=B(X)F_{\nu\alpha} F_{\mu}{^{\alpha}} + \frac{1}{2}g_{\mu\nu}A(X) \, .
    \label{tau2}
\end{equation}
Then, by comparing Eqs. \eqref{fieldeqsfull} and \eqref{tau2}, we identify
\begin{equation}
    A(X)=-\lambda h(T) \, ,
\end{equation}
and that the derivative of $A(X)$ with respect to $X$ is equal to $B(X)$ once more
\begin{equation}
     B(X) = A_X=-\lambda h_T T_X = \frac{\lambda}{2\pi} h_T \left(X \varphi_{XX}-\frac{D-4}{4}\varphi_X\right) \, .
\end{equation}
Thus, Eq. \eqref{fieldeqsfull} can be displayed as
\begin{equation}
    G_{\mu\nu} = \frac{\kappa^2}{4 \pi}\left[\left(\varphi_X -\frac{4\pi}{\kappa^2}A_X\right)F_{\nu\alpha}F_{\mu}{^{\alpha}}+\frac{1}{2}g_{\mu\nu}\left(\varphi-\frac{4\pi}{\kappa^2}A\right)\right] \, .
\end{equation}
Indeed, we reached the same conclusion: an $f(R,T)=R +\lambda h(T)$ gravity model coupled to NED can also be interpreted as GR coupled to a modified NED sector, which we explicitly computed.

\subsection{Invariant NEDs}
The nonlinear transformation induced in the effective matter sector in the general NED case studied above indicates that there might be $f(R,T)$ theories in which the transformed NED theory might remain dynamically invariant. In other words, there might be  NED models that remain insensitive to redefinitions of the type of Eq. \eqref{redef}. We now investigate this curious case of invariant NEDs, namely, NEDs that remain invariant under rescaling. In this context, the general rescaling equation is
\begin{equation}
    \varphi - \frac{4\pi}{\kappa^2}A=\sigma \varphi \, ,
\end{equation}
where $\sigma$ is a (constant) rescaling parameter. The equation above can also be rewritten in the form
\begin{equation}
    h(T)= \frac{\kappa^2}{4 \pi \lambda} \left(\sigma -1\right)\varphi \, ,
    \label{eq_scaling}
\end{equation}
so one can explore different scenarios depending on the explicit expression of the function $h(T)$. Notice here that $T=T^{\text{NED}}$. Next, we explore the linear and power-law cases.
\subsubsection{Linear Case: $h(T)=T$}
In this case, and after some algebra, Eq. \eqref{eq_scaling} becomes
\begin{equation}
    X\varphi_X = \left[\frac{D}{4}-\frac{\kappa^2}{2\lambda}\left(\sigma-1\right)\right] \varphi \equiv \gamma \varphi \, ,
\end{equation}
whose solution is 
\begin{equation}
    \varphi\left(X\right) = \left(\frac{X}{X_0}\right)^\gamma \varphi_0 \, .
\end{equation}

\subsubsection{Power-law Case: $h(T)=T^{\alpha}$}
Considering a power-law function of the trace, Eq. \eqref{eq_scaling} renders
\begin{equation}
    \frac{D}{4} \varphi - X\varphi_X = \left[\frac{\kappa^2}{4\pi \lambda}\left(\sigma-1\right)\right]^{\frac{1}{\alpha}}2\pi\varphi^{\frac{1}{\alpha}} \, ,
\end{equation}
which is a Bernoulli differential equation and admits an exact analytical solution of the form 
\begin{equation}
    \varphi\left(X\right) = \left(\frac{2\pi \lambda \alpha}{4(\alpha-1)+D\alpha}+ \left(\frac{X}{X_0}\right)^\frac{D\alpha}{4(\alpha-1)} \right)^\frac{\alpha}{\alpha-1} \, .
\end{equation}

The above two examples show that there exist NED theories whose dynamics is not affected by modifications of the gravitational Lagrangian of the type $R \to R+\lambda T^\alpha$ and, therefore, they yield exactly the same solutions as they would in GR.

\section{$f(R,T)=R+\lambda T$ Model Coupled to Scalar Matter}
\label{secIV}
Now, let us consider a linear $D$-dimensional $f(R,T)$ gravity model coupled to a general scalar field sector with action 
\begin{equation}
    S=\frac{1}{2\kappa^2}\int_{\Omega}\sqrt{-g}\left(R + \lambda T\right) d^Dx + \frac{1}{2}\int_\Omega\sqrt{-g} F(\phi,Z) d^Dx \, ,
\label{action_linear_scalar}
\end{equation}
where $F(Z,\phi)$ is a function of a scalar field $\phi$ and of a kinetic term $Z:=g^{\mu\nu}\partial_\mu \phi \partial_\nu \phi$. By varying Eq. \eqref{action_linear_scalar} with respect to the metric, we obtain the corresponding field equations
\begin{equation}
    G_{\mu\nu} + \lambda\left(\frac{\delta T^{\phi}}{\delta g^{\mu\nu}}-\frac{1}{2}g_{\mu\nu}T^{\phi}\right) = \kappa^2 T_{\mu\nu}^{\phi} \, ,
\label{field_eqs_raw_scalar}
\end{equation}
where the energy-momentum tensor $T_{\mu\nu}^{\phi}$ assumes the form
\begin{equation}
    T_{\mu\nu}^{\phi} = \frac{1}{2}g_{\mu\nu}F - F_Z \partial_\mu \phi \partial_\nu \phi \, ,
\end{equation}
where $F_Z$ denotes the partial derivative of $F$ with respect to the kinetic term $Z$. To check the conservation of $T_{\mu\nu}^{\phi}$, we proceed like in the NED case, considering first the GR scenario ($\lambda=0$). We need to use the corresponding equations of motion for the field, which for $\lambda=0$ take the form $F_\phi-2\nabla_\mu(F_Z \partial^\mu \phi)=0$, and note that $\nabla_\nu F=F_\phi \nabla_\nu \phi+F_Z \nabla_\nu Z$, with $\nabla_\nu Z=2 \partial^\mu \phi\nabla_\nu \partial_\mu\phi$. With this, it is immediately clear that 
\begin{equation}\label{eq:cons_scalar}
    \nabla_\mu {T^\mu}_\nu=-\nabla_\mu(F_Z\partial^\mu \phi)\partial_\nu \phi-F_Z \partial^\mu \phi \nabla_\mu \partial_\nu \phi+ \frac{1}{2}F_\phi \nabla_\nu\phi+ \frac{1}{2}F_Z\nabla_\nu Z
\end{equation}
vanishes when the equations of motion are satisfied. 

To check the conservation in the $\lambda\neq 0$ case, we proceed as before. First, we obtain the trace of the energy-momentum tensor, which is
\begin{equation}
   T^{\phi} = \frac{D}{2}F-ZF_Z \, ,
\end{equation}
and from this the following variation is verified
\begin{equation}
    \frac{\delta T^{\phi}}{\delta g^{\mu\nu}} =  \left(\frac{D-2}{2}F_Z + ZF_{ZZ}\right)\partial_\mu \phi \partial_\nu \phi\, .
\end{equation}
Taking into account these results, the explicit form of Eq. \eqref{field_eqs_raw_scalar} is
\begin{equation}
    G_{\mu \nu}+\lambda\left[\left(\frac{D-2}{2}F_Z - Z F_{ZZ}\right)\partial_\mu \phi \partial_\nu \phi -\frac{1}{2}g_{\mu\nu}\left(\frac{D}{2}F -ZF_Z\right)\right] =\kappa^2 \left(\frac{1}{2}g_{\mu\nu}F - F_Z \partial_\mu \phi \partial_\nu \phi\right) \, ,
\end{equation}
which can be equivalently rewritten as
\begin{equation}
    G_{\mu\nu}=\frac{1}{2}g_{\mu\nu}\left[\kappa^2 F+\lambda\left(\frac{D}{2}F-ZF_Z\right)\right]-\left[\kappa^2 F_Z +\lambda\left(\frac{D-2}{2}F_Z - Z F_{ZZ}\right)\right]\partial_\mu \phi \partial_\nu \phi \, .
\end{equation}
Once again, by defining a new scalar function $\mathcal{F}$ as
\begin{equation}
    \mathcal{F}:= F+ \frac{\lambda}{\kappa^2}\left(\frac{D}{2}F-ZF_Z\right) \, ,
\end{equation}
whose derivative with respect to $Z$ is
\begin{equation}
    \mathcal{F}_Z=F_Z+\frac{\lambda}{\kappa^2}\left(\frac{D-2}{2}F_Z -Z F_{ZZ}\right)\, , 
\end{equation}
subsequently, one writes the field equations of the theory as 
\begin{equation}
    G_{\mu\nu} = \kappa^2\left(\frac{1}{2}g_{\mu\nu}\mathcal{F}- \mathcal{F}_Z \partial_\mu \phi \partial_\nu \phi\right) \equiv \kappa^2 \Tilde{T}_{\mu\nu}^\phi  \, ,
\end{equation}
in which the right-hand side represents a modified scalar field source, which is analogous to the result found in the NED case. The same occurs if we have an $f(R,T)=R+\lambda h(T)$. In both cases, the resulting energy-momentum tensor is conserved by the same arguments as given in the discussion of Eq. (\ref{eq:cons_scalar}). The situation, however, changes if we combine sources of different types simultaneously because $h(T)$ models can lead to nonlinear mixings, as we are going to see next.

Before concluding this section, we note that, like in the NED case, one could consider the set of scalar field theories whose dynamics under an $f(R,T)=R+\lambda T^\alpha$ remain the same as in GR. Since the procedure to identify such a family is analogous to the NED case, we do not repeat that calculation here. 

\section{$f(R,T)=R+\lambda h(T)$ Model Coupled to Several Sources}
\label{secV}
As a final case, we consider the model $f(R,T)=R+\lambda h(T)$ coupled to a matter sector that is composed of matter coming from NED and scalar matter, i.e.
\begin{equation}
    S = \frac{1}{2 \kappa^2} \int_\Omega\sqrt{-g}\, \left[R + \lambda h(T)\right] \, d^D x + S^{\text{NED}} + S^{\phi} \, ,
    \label{action_total}
\end{equation}
where 
\begin{equation}
    S^{\text{NED}} = \frac{1}{8\pi}\int_\Omega \sqrt{-g} \varphi(X) d^Dx \, ; \, X := -\frac{1}{2}F_{\mu\nu}F^{\mu\nu} \, ,
\end{equation}
and 
\begin{equation}
    S^\phi = \frac{1}{2}\int_\Omega \sqrt{-g} F(Z,\phi) d^Dx \, ; \, Z := g^{\mu\nu} \partial_\mu \phi \partial_\nu \phi \, .
\end{equation}
Varying Eq. \eqref{action_total} with respect to the metric yields
\begin{equation}
    G_{\mu\nu} + \lambda\left(h_T \frac{\delta T}{\delta g^{\mu\nu}}-\frac{1}{2}g_{\mu\nu}h\right) = \kappa^2 T_{\mu\nu}=\kappa^2 \left(T_{\mu\nu}^{\text{NED}}+T_{\mu\nu}^{\phi}\right) \, ,
\end{equation}
where
\begin{equation}
T_{\mu\nu}^{\text{NED}} = \frac{1}{4\pi}\left(\varphi_X F_{\nu\alpha} F_{\mu}{^{\alpha}} +\frac{1}{2}g_{\mu\nu} \varphi\right) \, ,
    \label{energy-momentum tensor NED2}
\end{equation}
and 
\begin{equation}
    T_{\mu\nu}^{\phi} = \frac{1}{2}g_{\mu\nu}F - F_Z \partial_\mu \phi \partial_\nu \phi \, .
\end{equation}
Accordingly, the variation $\delta T/\delta g^{\mu\nu}$ is also constituted by two terms, which render
\begin{equation}
    \frac{\delta T}{\delta g^{\mu\nu}} = \frac{\delta T^{\text{NED}}}{\delta g^{\mu\nu}} + \frac{\delta T^{\phi}}{\delta g^{\mu\nu}} =\frac{1}{2\pi}\left(X\varphi_{XX}-\frac{D-4}{4}\varphi_X\right)F_{\nu\alpha} F_{\mu}{^{\alpha}} + \left( ZF_{ZZ}+\frac{D-2}{2}F_Z \right)\partial_\mu \phi \partial_\nu \phi\, .
    \end{equation}
Indeed, $\delta T/\delta g^{\mu \nu}$ is linear in $T^{\text{NED}} + T^{\phi}$. However, if $h(T)$ is a nonlinear function of $T$, then 
\begin{equation}
    h_T\frac{\delta T}{\delta g^{\mu\nu}} \neq h_{T^{\text{NED}}} \frac{\delta T^{\text{NED}}}{\delta g^{\mu\nu}} + h_{T^{\phi}} \frac{\delta T^{\phi}}{\delta g^{\mu\nu}} \, .
\end{equation}
As a result, there is a mixing between the terms from NED and the scalar field $\phi$ that prevents a clear splitting of the action terms in the same types of sources. In other words, there is a mixing between scalar and vector degrees of freedom, and it is not clear which field combinations will play the role of scalar and vector in the resulting effective theory. Nevertheless, one can always define an effective tensor $\tau_{\mu\nu}^{\text{Total}}$ that is conserved due to the Bianchi identities
\begin{equation}
\tau_{\mu\nu}^{\text{Total}} := \kappa^2 T_{\mu\nu} - \lambda\left(h_T \frac{\delta T}{\delta g^{\mu\nu}} - \frac{1}{2}g_{\mu\nu}h\right) \, .
\end{equation}
Yet, it is not clear what is conserved due to the exchanges between the scalar and NED fields.

\section{Summary and Conclusions}
\label{secVI}
In this work, we have explored some formal aspects of the dynamics of modified theories of gravity of the $f(R, T)$ type. Even in the absence of explicit couplings between the scalars $R$ and $T$, it is generally claimed that these theories lead to violations in the conservation of the energy-momentum tensor. However, we have shown that this fact is not always so. In particular, if one considers a single matter source of scalar or electromagnetic nature in theories of the form $f(R,T)=R+\lambda h(T)$, the effect of the modified dynamics is equivalent to a modification of the matter sector, which generically becomes nonlinear. We have shown explicitly how this occurs when an arbitrary NED Lagrangian $\varphi(X)$ or an arbitrary quintessence Lagrangian $F(\phi,Z)$  is taken, being the result valid in arbitrary dimension $D$. This also allowed us to see that there are some special families of theories for which the dynamics remain unchanged as compared to those found in GR. However, we also saw that when different kinds of matter are considered simultaneously, such as a NED plus a scalar sector, in nonlinear $f(R,T)=R+\lambda h(T)$ models, there is an explicit mixing of the elements of the matter sector that prevents their interpretation as the result of a (new and nonlinear) NED plus scalar sector in GR. This is so simply because the interactions induced between the matter fields via the $h(T)$ term (of gravitational origin) cannot be re-coded in self-interactions of the participating fields due to the nonlinear couplings that arise between the scalar and vector sectors of the matter terms. The results presented here are thus useful to better understand the nature of the mechanisms that induce new dynamics in this type of theories \cite{Pinto:2025loq}, where nonlinear self-interactions do not need to imply violations of energy-momentum conservation.

To conclude, one should note that for more general Lagrangians, not restricted to the decomposition $f(R,T)=R+\lambda h(T)$, the explicit non-minimal coupling between the geometric and matter sectors via products of curvature and matter, $R^n T^m$ (where $n$ and $m$ are real exponents), will in general lead to violations of energy-momentum conservation. Thus, in such a case, it is impossible to interpret the new matter terms, coming from $T^m$, as nonlinear realizations of the original matter action, as these terms will necessarily exchange energy and momentum with gravitons.

\section*{Acknowledgments}
MASP acknowledges support from the Funda\c{c}\~{a}o para a Ci\^{e}ncia e a Tecnologia, I.P. (FCT) research grants UIDB/04434/2020 (\url{https://doi.org/10.54499/UIDB/04434/2020}) and UIDP/04434/2020 (\url{https://doi.org/10.54499/UIDP/04434/2020}), and through the FCT project with reference PTDC/FIS-AST/0054/2021 (``BEYond LAmbda'') (\url{https://doi.org/10.54499/PTDC/FIS-AST/0054/2021}). MASP also acknowledges financial support from the FCT through the Fellowship UI/BD/154479/2022 (\url{https://doi.org/10.54499/UI/BD/154479/2022}). The authors also acknowledge financial support from the project i-COOPB23096 (funded by CSIC), and the Spanish Grants  PID2020-116567GB-C21 and PID2023-149560NB-C21, funded by MCIN/AEI
/10.13039/501100011033, and by CEX2023-001292-S funded by MCIU/AEI.  The paper is based upon work from COST Actions CosmoVerse CA21136 and CaLISTA CA21109, supported by COST (European Cooperation in Science and Technology).

\end{document}